\documentclass{webofc}
\usepackage[varg]{txfonts}   
\usepackage{wrapfig}
\usepackage{graphicx,caption}

\begin{document}
\title{Measurement of global polarization of $\Lambda$ hyperons in Au+Au $\sqrt{s_{\mathrm{NN}}}$ = 7.2 GeV fixed target collisions at RHIC-STAR experiment}
%
%

\author{\firstname{Kosuke} \lastname{Okubo (for the STAR Collaboration)}\thanks{\email{kokubo@rcf.rhic.bnl.gov}}
}

\institute{University of Tsukuba, 1-1-1 Tenno-dai, Tsukuba, Ibaraki 305-8571, Japan}

\abstract{%
  In non-central nuclear collisions, the created matter should exhibit strong vorticity, which induces spin polarization. 
Global polarization is a good tool to investigate the dynamics of the system. Global polarization of $\Lambda$ and $\bar{\Lambda}$ hyperons has been measured at $\sqrt{s_{\mathrm{NN}}}$ = 2.4 GeV - 5.02 TeV in various experiments. We have measured the global polarization of $\Lambda$ hyperons in Au+Au collisions with fixed-target configuration at $\sqrt{s_{\mathrm{NN}}}$ = 7.2 GeV at RHIC-STAR experiment.
The non-zero signal is observed and it follows the global trend of $\Lambda$ global polarization that increases towards lower collision energies. The global polarization is larger in more peripheral collisions but it does not show any significant rapidity dependence as well as transverse momentum dependence within uncertainties.
}

\maketitle
%
\section{Introduction}
\label{intro}
 In non-central heavy-ion collisions, the created matter should exhibit a rotational motion because of the conservation of the initial orbital angular momentum carried by the two nuclei, which is perpendicular to the reaction plane. Particle's and anti-particle's spins are aligned due to the spin-orbit coupling \cite{PH2005}.
The strong magnetic field would appear in the initial state and its direction is almost the same as the initial angular orbital momentum. Particle's and anti-particle's spins also could get aligned in the opposite direction by the magnetic field due to the opposite signs of their magnetic moments.
Therefore, global polarization is a good tool to investigate the dynamics of the system. Global polarization can be measured with hyperons through their parity-violating weak decays.
The STAR Collaboration reported the global polarization of $\Lambda$ and $\bar{\Lambda}$ hyperons at $\sqrt{s_{\mathrm{NN}}}$ = 7.7 - 200 GeV and observed positive signals \cite{PH2007,PHnature,PHniida}.
Additionally, the ALICE Collaboration measured global polarization of $\Lambda$ and $\bar{\Lambda}$ hyperons at $\sqrt{s_{\mathrm{NN}}}$ = 2.76, 5.02 TeV \cite{PHalice}.
These measurements show that global polarization becomes larger at lower collision energies.
 The STAR fixed-target program provides an opportunity to extend such measurements at even lower energies.
We here present the measurement of global polarization of $\Lambda$ hyperons in Au+Au collisions at $\sqrt{s_{\mathrm{NN}}}$ = 7.2 GeV with fixed-target configuration.

\section{Analysis details}
 The good minimum bias data of Au+Au collisions have been collected by the STAR detector with fixed-target configuration in 2018 and 209 million events were analyzed in this study. The gold target has been installed inside the vacuum pipe at 2.0 meters away from the center of the STAR detector.
The decay channel of $\Lambda \rightarrow p + \pi$ (BR: 63.9\% \cite{PDG}) was utilized for $\Lambda$ identification. The daughter particles were identified using TPC d$E$/dx and TOF and then $\Lambda$ hyperons were reconstructed based on the invariant mass, applying cuts on decay topology.
In this parity-violating weak decay, the daughter proton preferentially decays along the $\Lambda$'s spin direction.
Furthermore, since the system angular momentum is perpendicular to the reaction plane, the global polarization can be measured via the distribution of the azimuthal angle of the daughter proton in the $\Lambda$'s rest frame with respect to the reaction plane as follows \cite{PH2007}:
\begin{equation}
\label{eq1}
P_H = \frac{8}{\pi\alpha_H}\frac{\langle\sin(\Psi_1-\phi^*_p)\rangle}{\mathrm{Res}(\Psi_1)}
\end{equation}
where $\alpha_H$ is $\Lambda$'s decay parameter ($\alpha_{\Lambda} = 0.732\pm0.014$ \cite{PDG}), $\Psi_1$ is the first order event plane, and $\phi^*_p$ is the azimuthal angle of the daughter proton in the $\Lambda$'s rest frame.
The event plane reconstructed by Event Plane Detector (EPD) and the resolution estimated by three subevent method \cite{EPreso}. 
\begin{figure}[h]
\begin{center}
\includegraphics[width=55mm]{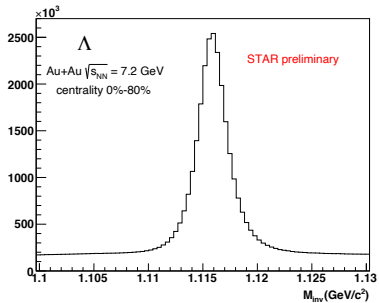}
\caption{Invariant mass distribution of $\Lambda$ hyperons in 0\%-80\% centrality for Au+Au collisions at $\sqrt{s_{\mathrm{NN}}}$ = 7.2 GeV.}
\label{fig:InvM}
\end{center}
\end{figure}

\section{Result}
\begin{figure}[h]
\begin{center}
\includegraphics[width=70mm]{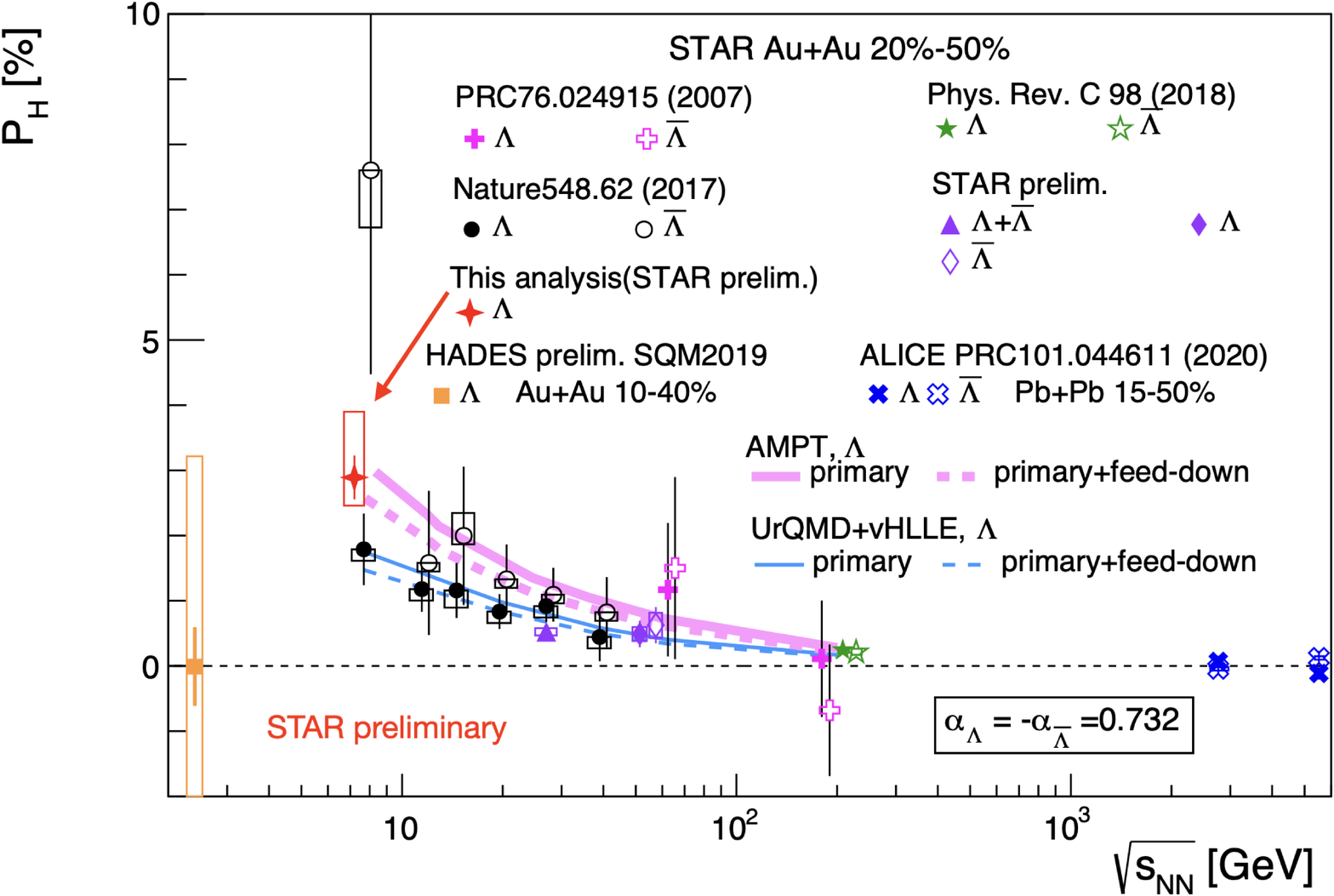}
\caption{Global polarization of $\Lambda$ (solid markers) and $\bar{\Lambda}$ (open markers) as a function of collision energy. Bold lines show the AMPT model calculations \cite{PHampt} and thin lines show calculations by a 3+1D cascade + viscous hydrodynamic model (UrQMD + vHELL) \cite{PHurqmd}. Primary $\Lambda$ with and without the feed-down effect are indicated by dashed and solid lines, respectively.}
\label{fig:PHvsEnergy}
\end{center}
\end{figure}

\begin{figure}[h]
\begin{center}
\includegraphics[width=65mm]{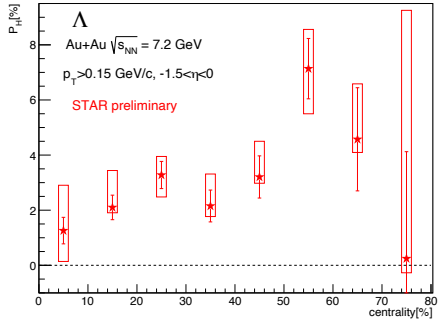}
\caption{Global polarization of $\Lambda$ hyperons as a function of centrality for $p_{\mathrm{T}}$ > 0.15 GeV/$c$, -1.5 < $\eta$ < 0.}
\label{fig:PHvsCent}
\end{center}
\end{figure}

 Figure \ref{fig:PHvsEnergy} shows global polarization of $\Lambda$ and $\bar{\Lambda}$ as a function of collision energy. Solid markers indicate $\Lambda$ and open markers indicate $\bar{\Lambda}$ global polarization.   
Because the $\Lambda$ and $\bar{\Lambda}$'s decay parameter were updated recently, the previous results were rescaled by using the new values ($\alpha_{\mathrm{old}}/\alpha_{\mathrm{new}}$).
The result at 7.2 GeV shows a non-zero positive signal. It follows the global trend of $\Lambda$ global polarization that increases towards lower collision energies.
\newpage
\noindent
We have also performed differential measurements such as centrality, rapidity, and transverse momentum dependence of global polarization.
Figure \ref{fig:PHvsCent} shows centrality dependence of $\Lambda$ global polarization.
$\Lambda$ global polarization increases in more peripheral collisions as expected from the initial collision geometry.
\par
\begin{figure}[htbp]
\begin{minipage}{0.5\hsize}
\begin{center}
\vspace{44mm}
\captionsetup{width=.85\linewidth}
\mbox{\raisebox{5mm}[-3mm][0pt]{\includegraphics[width=60mm]{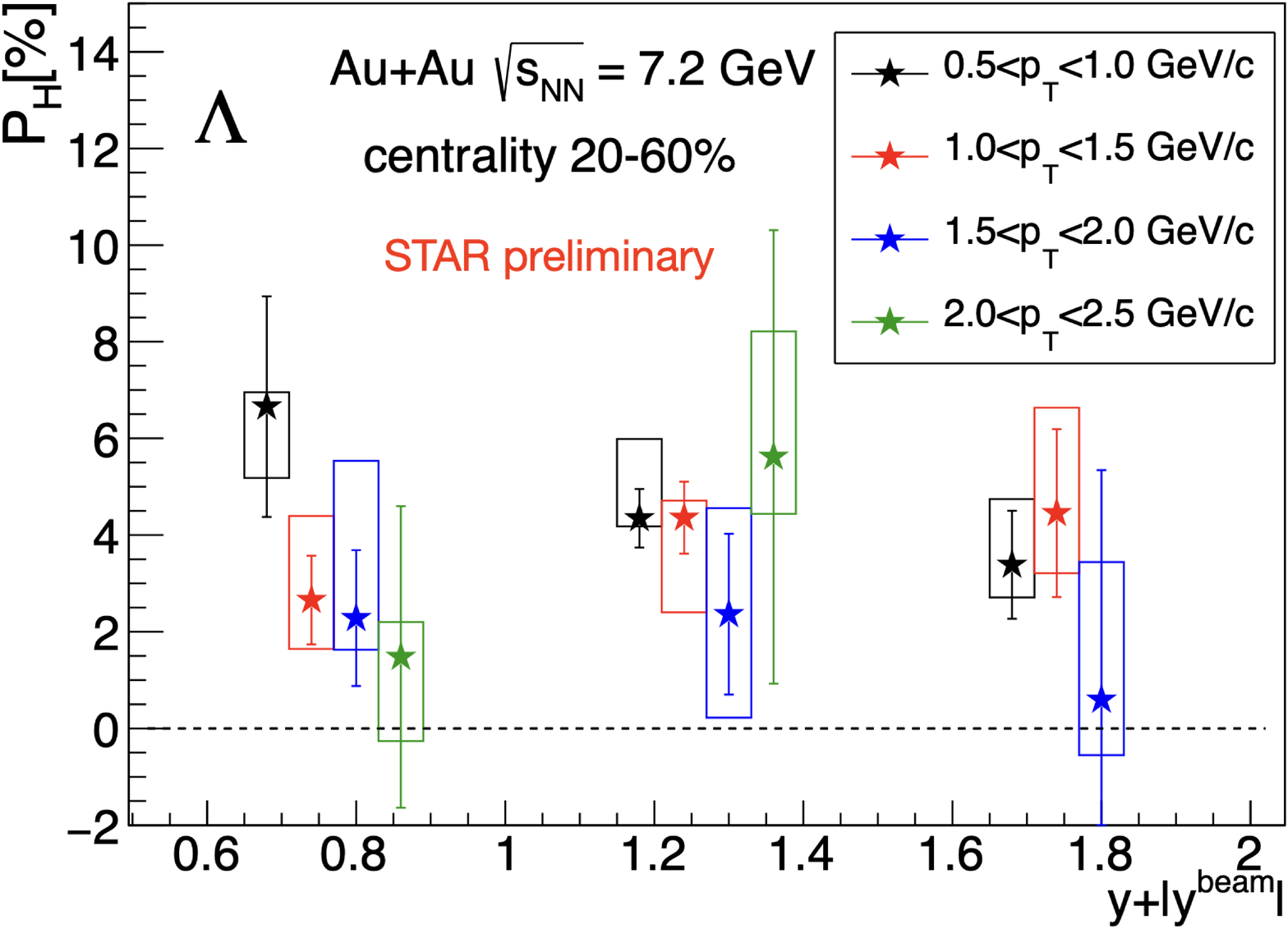}}}
\vspace{-5mm}
\caption{Global polarization of $\Lambda$ hyperons  as a function of rapidity for 20\%-60\% centrality each $p_{\mathrm{T}}$ region.}
\label{fig:PHvsRapidity}
\end{center}
\end{minipage}
\begin{minipage}{0.5\hsize}
\begin{center}
\captionsetup{width=.85\linewidth}
\includegraphics[width=60mm]{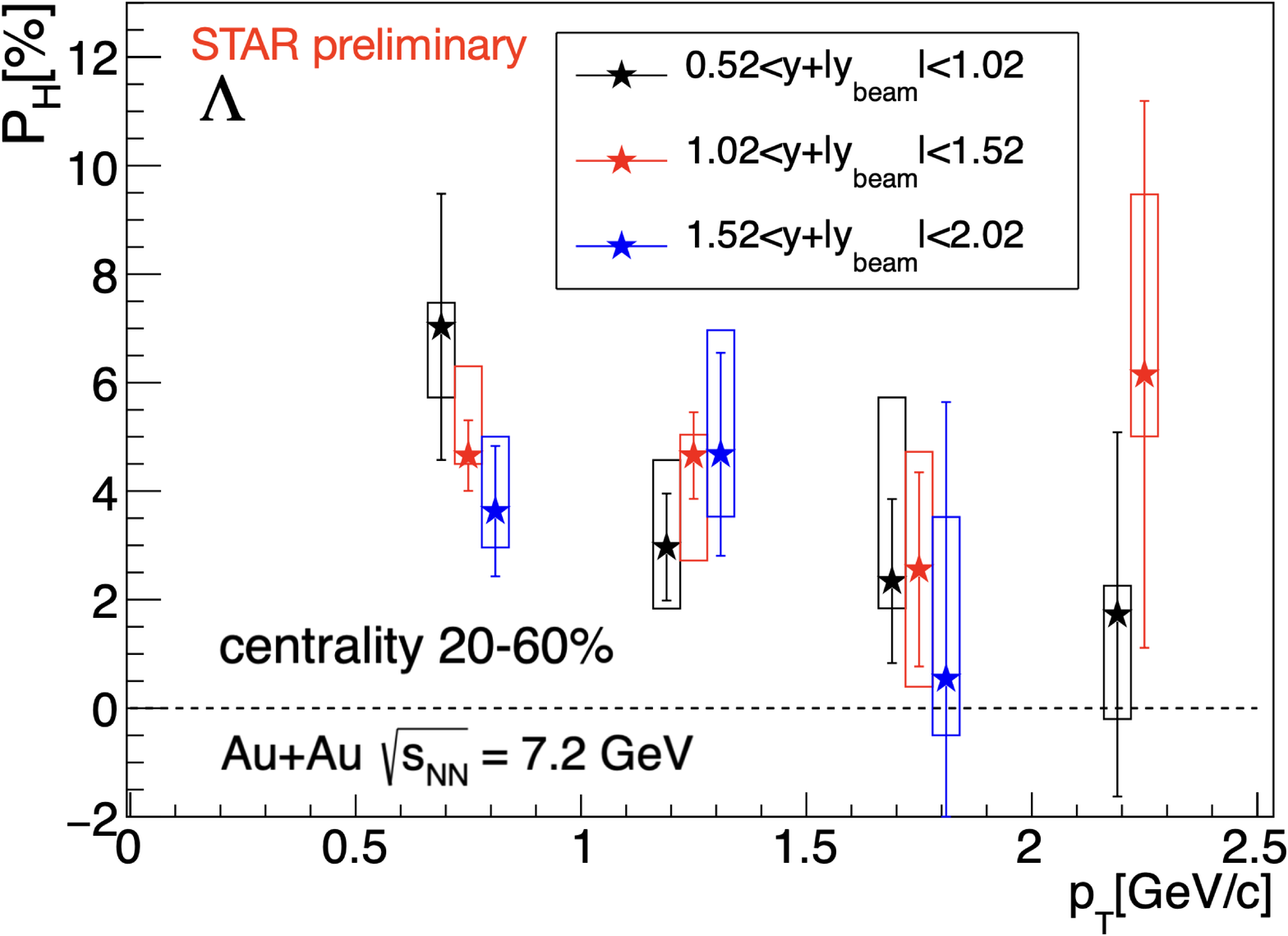}
\caption{Global polarization of $\Lambda$ hyperons as a function of transverse momentum for 20\%-60\% centrality each rapidity region.}
\label{fig:PHvsPt}
\end{center}
\end{minipage}
\end{figure}

\noindent
Theoretical models predict that the polarization depends on rapidity, although the prediction is different among the models \cite{PHrapiHIJING,PHrapiGeometry,PHrapiAMPT,PHrapiCLVisc}.
Figure \ref{fig:PHvsRapidity} shows rapidity dependence of $\Lambda$ global polarization for 20\%-60\% centrality, shown for 4 different $p_{\mathrm{T}}$ regions.
$\Lambda$ global polarization is consistent with being constant within uncertainties within the  acceptance of the STAR detector. STAR experiment has recently installed a new detector inner Time Projection Chamber (iTPC) and will upgrade the forward detectors (2023+2025). With these upgrades, global polarization in large rapidity regions can be explored in the future.
The $\Lambda$ global polarization is expected to decrease at low $p_{\mathrm{T}}$ due to the smearing effect caused by scattering at the later stage of the collision. On the other hand, it may also decrease at high $p_{\mathrm{T}}$ due to the jet fragmentation.
Figure \ref{fig:PHvsPt} shows $\Lambda$ global polarization as a function of transverse momentum for 20\%-60\% centrality. 
The $\Lambda$ global polarization does not show significant $p_{\mathrm{T}}$ dependence within uncertainties.

\section{Summary}
We present the first measurement of $\Lambda$ global polarization in Au+Au collisions at $\sqrt{s_{\mathrm{NN}}}$ = 7.2 GeV with STAR fixed-target configuration. The positive non-zero signal has been observed and it follows the global trend of energy dependence. We have also performed differential measurements of global polarization.
The increasing trend of polarization towards peripheral collisions is observed. We have observed no significant rapidity and $p_{\mathrm{T}}$ dependence within uncertainties. With newly upgraded iTPC and future upgrade of forward detectors in 2023-2025, the global polarization will be explored in more forward regions.

%
%
%

\end{document}